# Design and Finite-Element Analysis of a New Inclined Blade-Electrode Architecture for Trapped-Ion Quantum Information Processing

Nahiyan Archa[1], Abhinand P[1], Ahammed Shabeeb[2], Nikhil Kumar[3,*]

[1] Department of Chemistry, National Institute of Technology Calicut, India.
[2] Department of Physics, National Institute of Technology Calicut, India.
[3] Department of Electronics and Communication Engineering, National Institute of Technology Calicut, India.
* Author to whom any correspondence should be addressed.

**Email:** nikhilkumarcs@nitc.ac.in

**Abstract**

The development of scalable quantum technologies depends on ion-trap architectures that deliver strong, stable confinement while minimizing motional heating from electric-field noise. Trapped-ion systems exhibit exceptionally long coherence times and high-fidelity quantum control; however, anomalous motional heating from electrode surfaces can constrain performance. This study presents a finite-element computational analysis of a novel inclined ion-trap architecture to examine the interplay among electrode geometry, inter blade distance, confinement strength, secular frequencies, motional heating, thermal response, and stability analysis. Five trap inclinations were evaluated using three-dimensional electrostatic simulations for blade separations ranging from 5 to 35 μm. The RF electric-field distribution was used to calculate the effective pseudopotential, trap depth, secular frequencies, and corresponding normal modes. An empirical electric-field-noise model was then applied to estimate the motional heating rate for frequency-noise exponents α = 3, 3.5, and 4. The findings indicate a pronounced trade-off between confinement strength and ion–electrode distance, with intermediate electrode separations achieving a favorable balance between trap depth, secular frequencies, and predicted motional heating. The analysis identifies the blade-separation range of approximately 25–35 μm as the key design regime, where decreasing the separation provides stronger confinement and lower predicted heating but reduces the RF stability margin, while increasing the separation improves stability at the expense of weaker confinement and higher predicted heating. Near 25 μm, the heating is lower about 127–135 quanta $s^{-1}$ at α = 3 but the RF stability is lower ($q_{max} \approx 1.9$), while near 35 μm the stability is much better ($q_{max} \approx 0.52$) but the heating increases to about 1570–1820 quanta $s^{-1}$ at α = 3. These results establish a computational framework for the design and optimization of inclined traps and offer design guidelines for future trapped-ion quantum-information architectures.

## 1. Introduction

Gauss's law ($\nabla \cdot E = 0$) makes it practically impossible to confine charged particles in three dimensions using just static electric fields, but a time-varying and spatially inhomogeneous electric field can produce a time-averaged restoring force for trapped particles, which is the basic idea behind the Paul trap [1]. If trapped in an ultra-high vacuum environment, these atomic ions can be isolated from their surroundings and manipulated individually [2]. An ideal microwave near-field design at the trapped-ion spot [3], or a concentrated laser with beam-steering capacity, can be used to perform these manipulations [4]. Theoretically, spontaneous emission-based million-year coherence is supported by trapped atomic ions [5-7], but when it comes to experimentation, we have $(3.77 \pm 1.09) \times 10^4$ seconds, which is more than ten hours and breaks the record [8]. These characteristics make the ion trap approach a viable option for putting quantum information processing into practice [9] and for accurate measurement in basic physics applications such as mass spectrometry and optical clocks [10-12].

In the initial ion-trap quantum computation idea, a string of ions was contained in a single trap; their electronic states served as qubit logic levels, and quantum information was transferred between the ions via their mutual Coulomb interaction [13]. Because individual motional modes and ions are harder to control as the number of ions in a trap increases, researchers designed large-scale ion-trap quantum computers by connecting segments of smaller ion traps [14]. Although it was well received, it faced challenges such as anomalous motional heating [15], fast ion shuttling with minimal decoherence [16, 17], and demonstrating multi-ion manipulation in segmented traps [18]. Microfabrication technologies are a preferred approach for creating different ion traps [19], but they suffer from heating from various noise sources, which can degrade coherence and fidelity [20].

Here we present a new architecture, see Fig. 1, which reduces trap heating. The idea grounds in increasing the blade-ion distance as much as possible; this is acquired by exploiting the vector space, as in Fig. 2. Which helps us to create a pseudopotential curvature which can trap the ions. As the θ increases, the distance between the ion and the plane in which the trap lies increases and the confinement through the x- and y-axes decreases while increasing it along the z-axis.

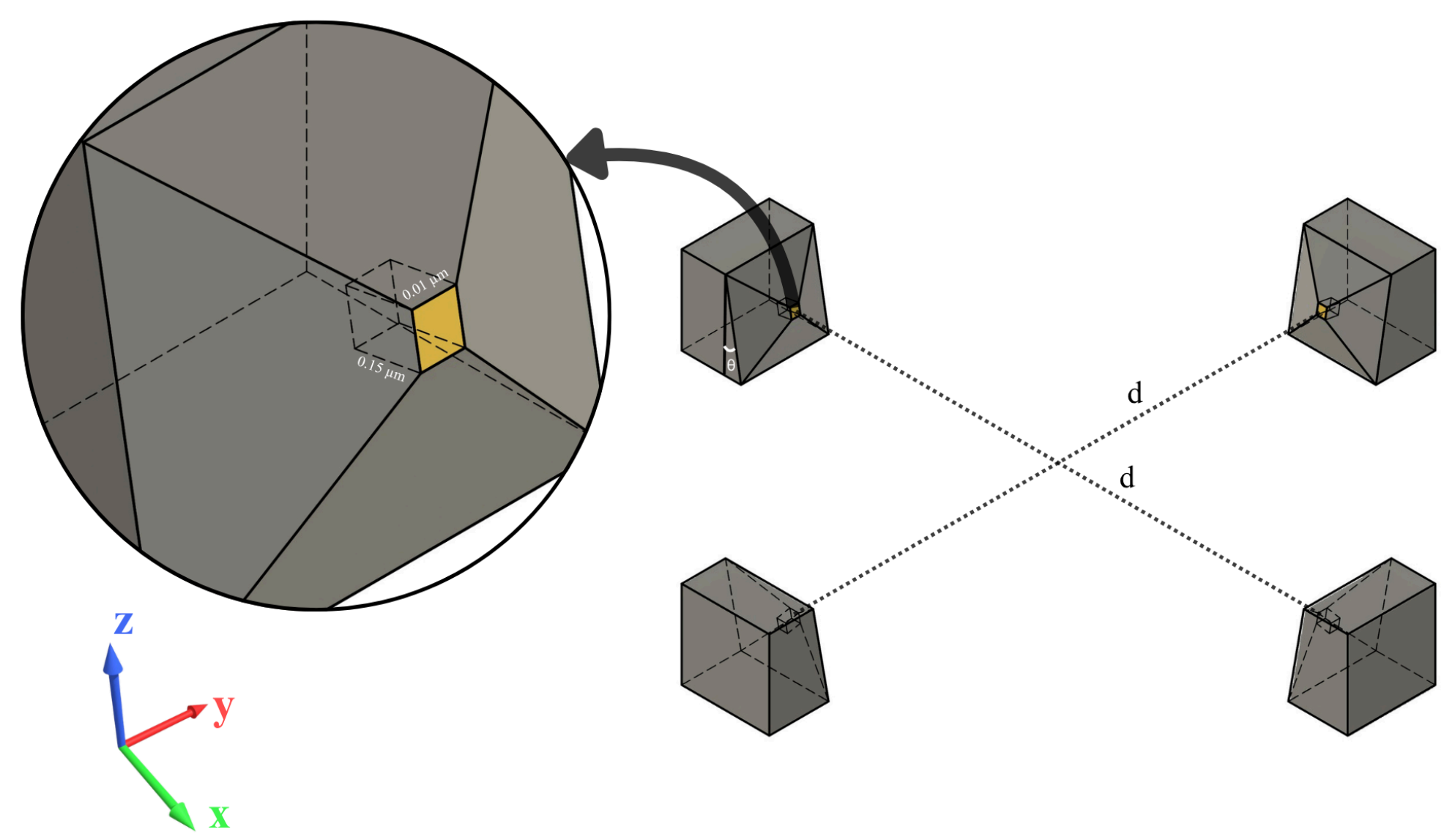

Figure 1. Schematic of the inclined blade-electrode trap geometry. Four blade electrodes are arranged symmetrically about the trap centre, each inclined at angle θ and positioned at distance d from each other. Inset: magnified view of the gold electrode (0.15 μm × 0.01 μm × 0.01 μm), the feature size used in the finite-element model. Coordinate axes (x, y, z) indicate the frame used throughout the paper.

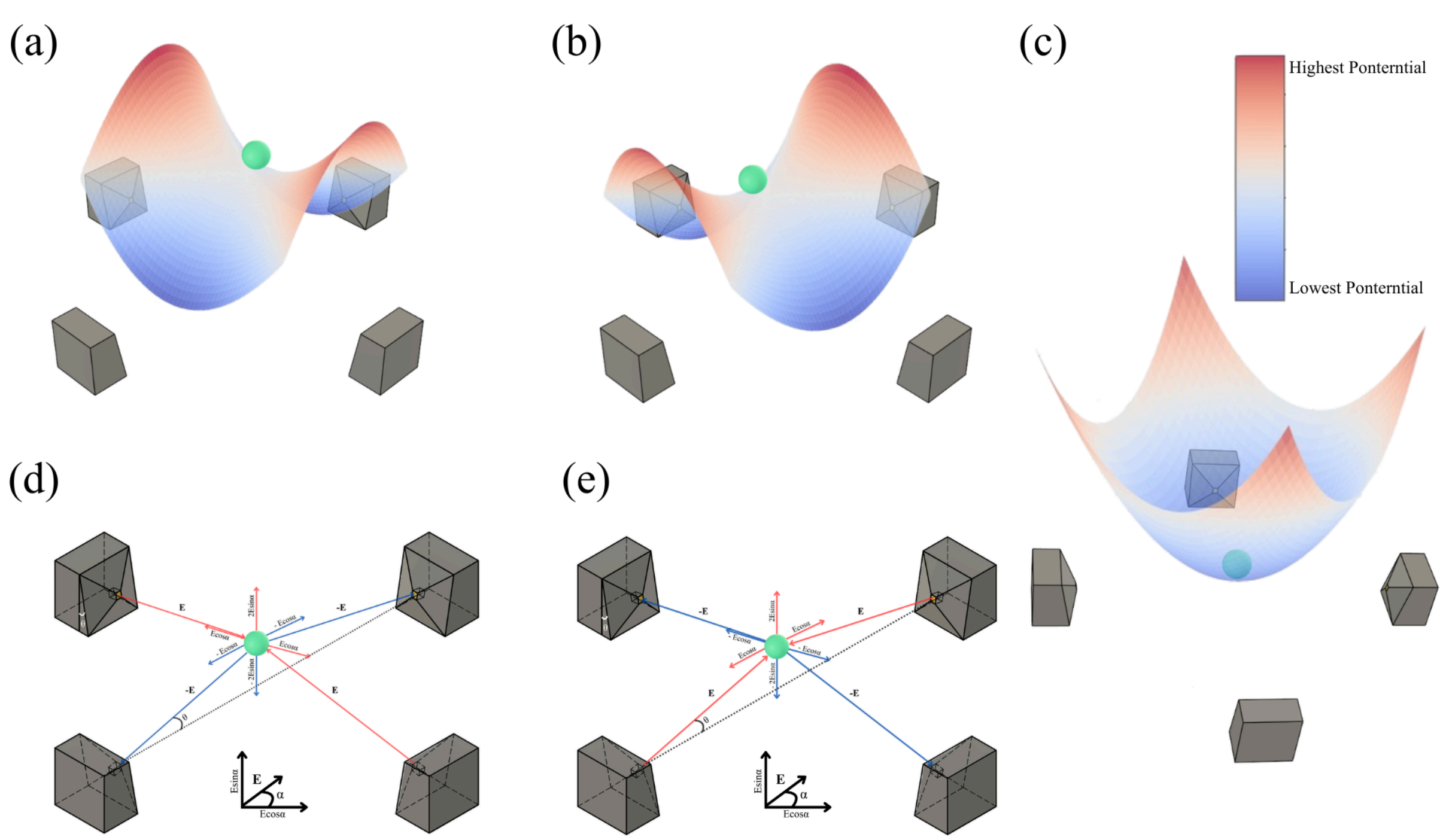

Figure 2. Pseudopotential confinement and RF field configuration for the inclined trap. (a) 3D potential curve at instant t = 1, (b) t = 2, (c) 3D pseudopotential curve at respective frequency, (d) electric field (**E**) in 3D vector space at instant t = 1 and (e) t = 2.

## 2. Methods

### *2.1. Trap geometry and finite-element model*

Three-dimensional electrode geometries for the proposed ion traps were created using Autodesk Inc. 2021 Fusion 360 [21]. Five inclined trap configurations, with inclinations ranging from 2° to 6°, were developed to examine the influence of electrode geometry and separation on trapping characteristics. The resulting CAD models were exported in STEP format and subsequently imported into COMSOL Multiphysics® v. 6.0 [22] for finite-element simulations. The imported geometries were carefully inspected and verified prior to the assignment of material properties, physics interfaces, and boundary conditions.

The characteristic dimensions of the electrode structures were defined on the micrometre scale. We varied the blade separation d from 5 to 35 μm. The trap was defined at (0,0,1) μm in the simulation coordinate system. The electrode structures were assigned gold material properties, while everything else was treated as vacuum.

### *2.2. Electrostatic field calculation*

In order to create the pseudopotential, we need to alternate the signs of electrodes as in (d) and (e) of Fig. 2 at enough frequency to trap the ion. Since the electrodes are inclined at a degree θ it will split the electric field (**E**) in the vector space where $2E\sin\alpha$ from the positive electrode cancels the $-2E\sin\alpha$ from the negative electrode, constraining the trap in the z-axis throughout all the instances. At t = 1, $E\cos\alpha$ cancels out each other along the x-axis and the $-E\cos\alpha$ along the y-axis, constraining the ion in the x- and y-axes respectively. This happens vice versa at t = 2. This repeated process creates a pseudopotential similar to the one in (c) of Fig. 2 trapping the ion at the trap sight.

The spatial electric-potential distribution generated by the RF electrode geometry was calculated using the Electrostatics (es) physics interface in COMSOL Multiphysics 6.0. A potential amplitude of [23]

$$V_{RF} = 15\ kV$$

was applied to the RF electrodes, with the opposing RF electrodes assigned the same potentials as to act as pseudopotential. The electrostatic solution provides the spatial distribution of the RF potential amplitude, ΦRF (r), and the corresponding electric-field amplitude was obtained from

$$E_{RF}(\mathbf{r}) = -\ \nabla\Phi_{RF}(\mathbf{r}).$$

The resulting electric-field magnitude and its spatial derivatives were evaluated at the trapping position for each electrode geometry and blade separation.

No time-dependent RF waveform was explicitly solved in COMSOL. Instead, the electrostatic solution was treated as the spatial RF-potential amplitude and subsequently used to calculate the effective RF pseudopotential. This was sufficient because the pseudopotential is determined by the spatial RF electric-field amplitude, while the temporal RF dependence is accounted for analytically through the driving frequency.

*2.3. RF pseudopotential and trap depth*

The effective pseudopotential for a singly charged $^{171}Yb^{+}$ ion was calculated from the electrostatic RF-field amplitude obtained from COMSOL [24, 25]:

$$\Psi(\mathbf{r}) = \frac{e^{2}|\mathbf{E}_{\mathbf{RF}}(\mathbf{r})|^{2}}{4m\Omega^{2}},$$

where e is the elementary charge, m is the mass of the $^{171}Yb^{+}$ ion, and

$$\Omega = 2\pi f_{RF}$$

is the RF angular frequency. A drive frequency of $f_{RF}$ = 28 MHz was used in the pseudopotential calculation.

The trap depth was determined from the difference between the pseudopotential at the relevant saddle point and that at the local minimum corresponding to the trapping position,

$$U_{depth} = \Psi_{saddle} - \Psi_{min}.$$

This procedure was repeated for all five electrode geometries and for blade separations between 5 and 35 μm.

*2.4. Secular frequencies and normal modes*

The local curvature of the calculated pseudopotential was described by its Hessian matrix,

$$H_{\Psi} = \nabla\nabla\Psi.$$

The three eigenvalues of this matrix were used to determine the principal curvatures of the trapping potential [26]. The corresponding secular frequencies were calculated as

$$f_i = \frac{1}{2\pi}\sqrt{\frac{\lambda_i}{m}},$$

where $\lambda_i$ denotes the i-th eigenvalue of the pseudopotential Hessian.

The corresponding eigenvectors define the principal motional directions of the trapped ion. This treatment retains the coupling between the Cartesian coordinates arising from the nonzero off-diagonal Hessian terms [26] and therefore does not require the proposed geometry to be approximated as an ideal quadrupole trap.

## 3. Results

### *3.1. RF pseudopotential confinement and trap depth*

Fig. 3(a) depicts the calculated trap depth as a function of blade separation d for the five inclination angles ($\theta$ = 2°-6°). The curves show a similar decreasing trend for all five geometries, with a near-complete overlap over the investigated range. This indicates that the blade separation is the dominant geometric parameter controlling the trap depth, while the small variation in inclination angle has only a weak effect on its overall magnitude. At $\theta$ = 2°, for example, the calculated trap depth decreases from approximately $7.2 \times 10^{-17}$ J at d = 5 μm to $5.2 \times 10^{-21}$ J at d = 35 μm. Similar behaviour is observed for the other inclination angles. The decrease in trap depth with increasing blade separation results from the reduction in the RF electric-field gradient at the ion position as the electrodes are moved farther apart. Since the effective RF pseudopotential is proportional to the square of the RF electric-field amplitude, a weaker field and smaller field curvature produce a shallower trapping potential. At d = 20 μm, the trap depths remain on the order of $10^{-19}$ J, while at d = 30 and 35 μm they decrease to the order of $10^{-20}$ and $10^{-21}$ J, respectively.

Fig. 3(b) shows a representative pseudopotential profile along the z-axis for d = 30 μm and $\theta$ = 6°. The profile exhibits a well-defined minimum near the trap centre, surrounded by higher-pseudopotential regions on both sides. This minimum corresponds to the equilibrium position of the ion, where the RF electric-field magnitude is minimized. The increase in pseudopotential away from this position forms the confining potential, while the lower of the surrounding potential barriers determines the trap depth. The approximately symmetric profile about the trap centre indicates that the proposed electrode arrangement produces confinement around a central equilibrium position along the investigated direction.

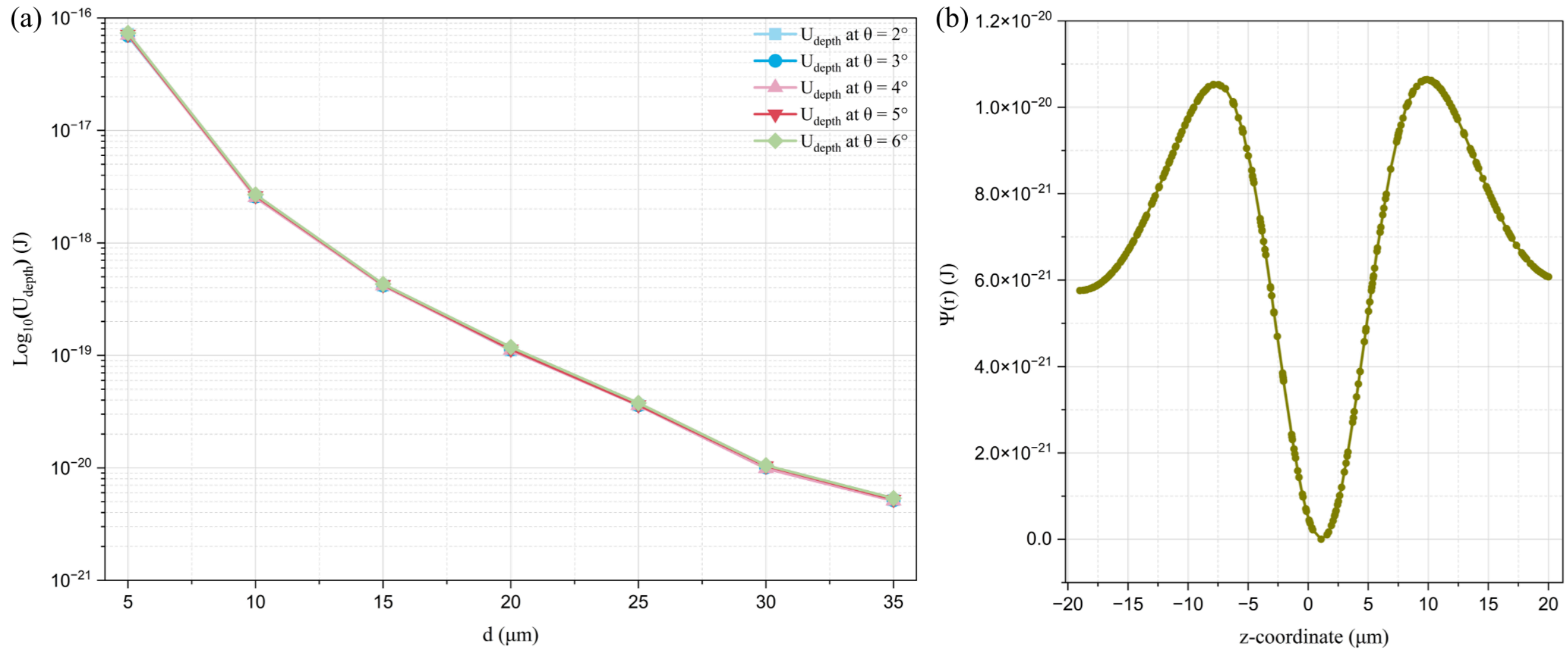

Figure 3. (a) Trap depth as a function of blade separation d for all five inclination angles (θ = 2°–6°), plotted on a logarithmic scale. The near-complete overlap of the five curves indicates that trap depth is governed primarily by d, with only a weak dependence on θ. (b) Representative pseudopotential Ψ(r) along the z-axis for d = 30 μm, θ = 6°, showing the confining minimum at the trap centre flanked by potential barriers whose lower value defines the trap depth $U_{depth}$.

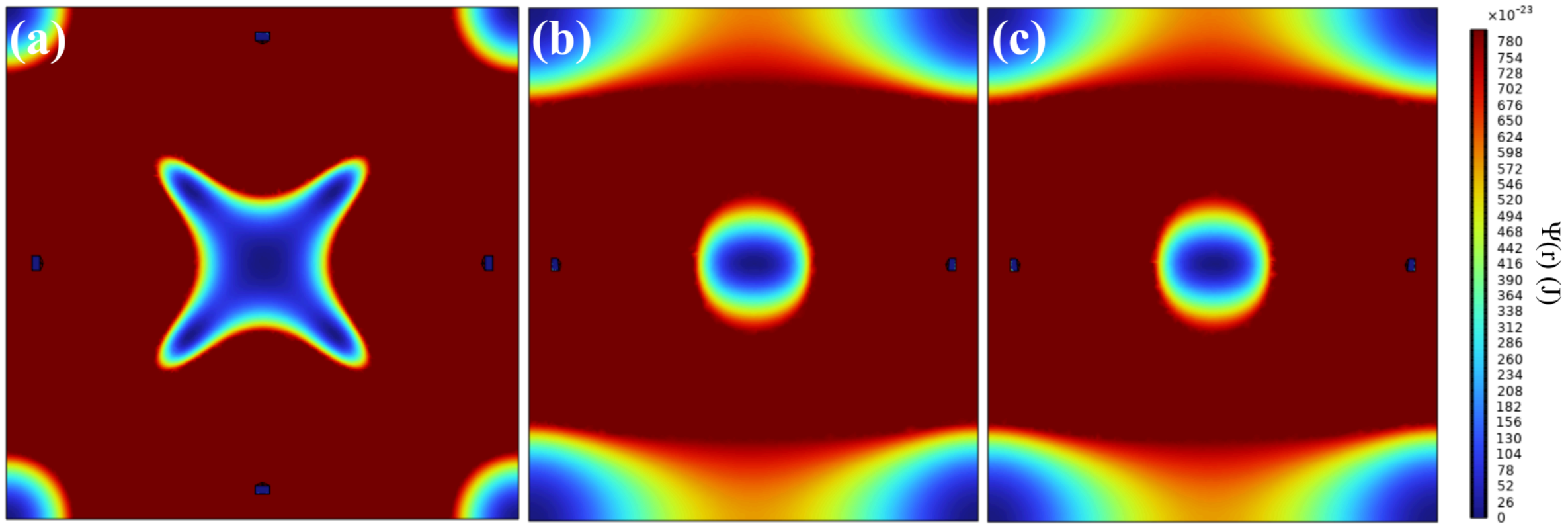

Figure 4. Pseudopotential contours for the proposed inclined electrode geometry at d = 30 μm and θ = 6° (a) xy-plane, (b) xz-plane, and (c) yz-plane. The localized low-pseudopotential region at the centre in all three planes indicates the formation of a three-dimensional RF trapping region.

The three-dimensional nature of this confinement is further illustrated in Fig. 4, which presents pseudopotential contours in the xy, xz, and yz planes for d = 30 μm and θ = 6°. The xy-plane shows a localized low-pseudopotential region at the centre, surrounded by higher-pseudopotential regions produced by the inclined blade electrodes. The corresponding xz- and yz-plane contours also show a localized minimum around the same central position, demonstrating that the confinement is not restricted to a single transverse plane but extends in all three spatial directions. The presence of a common central minimum in all three planes confirms the formation of a three-dimensional RF trapping region.

*3.2. Secular frequencies and normal modes*

Fig. 5 shows the variation of the three secular frequencies with blade separation dd for the five inclination angles, θ = 2°-6°. The three curves in each panel correspond to the three coupled normal modes, $f_1$, $f_2$, and $f_3$. A strong decrease in all three frequencies is observed as the blade separation increases. The largest change occurs between 55 and 10 μm, after which the frequencies decrease more gradually. The similar trends observed for all five inclination angles indicate that the blade separation has a stronger influence on the secular frequencies than the small variation in inclination angle.

At θ = 6°, for example, the calculated frequencies decrease from approximately 1377, 3665, and 4534 MHz at d = 5 μm to approximately 193, 254, and 426 MHz at d = 10 μm. With further increase in separation, the frequencies decrease to approximately 53, 55, and 107 MHz at d = 15 μm, and to approximately 20, 21, and 41 MHz at d = 20 μm. At the largest separation investigated, d = 35 μm, the frequencies reach approximately 2.50, 2.68, and 5.21 MHz. Thus, increasing the blade separation progressively weakens the local curvature of the pseudopotential and consequently reduces the motional frequencies.

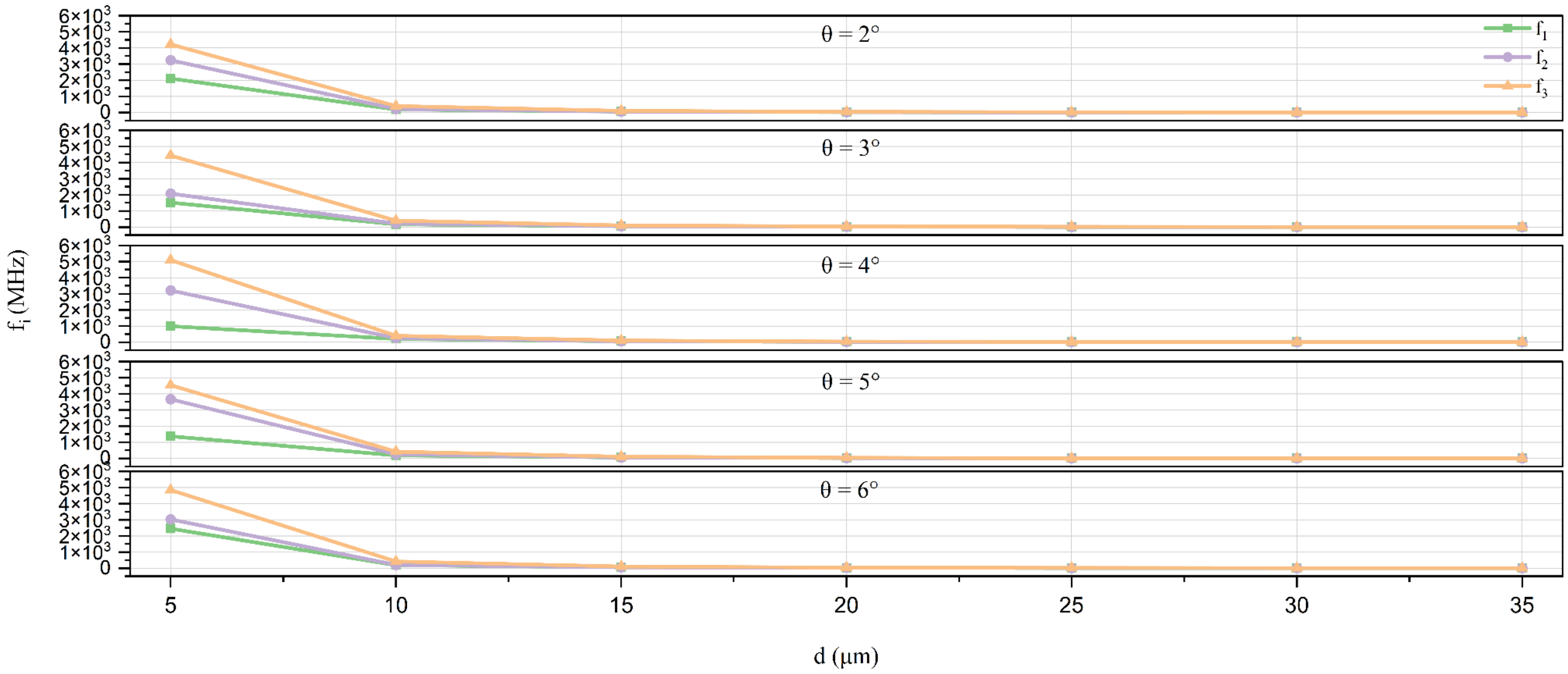

Figure 5. Secular frequencies of the three normal modes as a function of blade separation d for the five electrode inclination angles, θ = 2°-6°. The three curves in each panel correspond to the normal modes $f_1$, $f_2$, and $f_3$.

The three secular frequencies were obtained from the eigenvalues of the Hessian matrix of the calculated pseudopotential at the trapping position. Because the off-diagonal elements of the Hessian are generally nonzero for the inclined and non-planar electrode geometry, the motional coordinates are coupled and the frequencies were therefore treated as normal-mode frequencies rather than as independent Cartesian oscillations. The eigenvalues represent the principal curvatures of the pseudopotential well along the three normal-mode directions. Consequently, the strong reduction in frequency with increasing dd reflects the reduction in pseudopotential curvature as the ion is positioned farther from the electrodes.

The relative separation of the three modes also changes with blade separation. At d = 5–10 μm, the three modes are strongly non-degenerate, indicating pronounced anisotropy of the local confinement produced by the nearby electrode surfaces. For remaining electrode separations, the two lower-frequency modes become closely spaced, while the third mode remains approximately twice their average frequency across the five geometries. This behaviour corresponds approximately to a 1:1:4 ratio in the principal curvatures of the pseudopotential, since the secular frequency is proportional to the square root of the corresponding curvature. The convergence of the two lower-frequency modes therefore indicates that the confinement becomes increasingly similar along two normal-mode directions as the ion–electrode distance increases, whereas the third direction retains stronger confinement.

### *3.3. Electric-field noise and motional heating*

Fig. 6 shows the calculated motional heating rates as a function of blade separation d for the five inclination angles, θ = 2°-6°, considering the three frequency exponents α = 3, 3.5, and 4. Each panel corresponds to one inclination angle, while the different curves represent the three normal modes and the three assumed values of α. For all five geometries, the heating rate increases with increasing blade separation. The curves remain relatively close at small separations but diverge increasingly at larger separations, with the highest heating rates obtained for α = 4. This behaviour becomes particularly pronounced at d = 30–35 μm.

The predicted heating rate was estimated using the empirical electric-field-noise model

$$S_E(f, d) = S_0\left(\frac{d_0}{d}\right)^4\left(\frac{f_0}{f}\right)^\alpha \text{ [27]},$$

where $S_0 = 7.0 \times 10^{-13}$ $V^2m^{-2}Hz^{-1}$, $d_0 = 59$ μm, and $f_0 = 4.7$ MHz [28]. The three values α = 3, 3.5, and 4 were considered to represent the uncertainty in the frequency dependence of surface-related electric-field noise. The corresponding motional heating rate was calculated from

$$\dot{n} = \frac{e^2 S_E(\omega)}{4m\hbar\omega} \text{ [29]},$$

where m is the mass of the trapped $^{171}Yb^{+}$ ion and $\omega = 2\pi f$.

The increase in heating rate with blade separation can be understood from the combined dependence of the noise spectrum and the secular frequency. As the blades are moved farther apart, the ion-electrode distance increases, which reduces the distance-dependent electric-field noise according to the $d^{-4}$ term. However, the same increase in separation substantially reduces the secular frequencies, as shown in Fig. 5. Since the heating rate is inversely proportional to ω, this reduction in frequency increases the heating rate and can dominate over the reduction in the distance-dependent noise. Thus, increasing the ion–electrode distance does not necessarily result in lower motional heating in the present system.

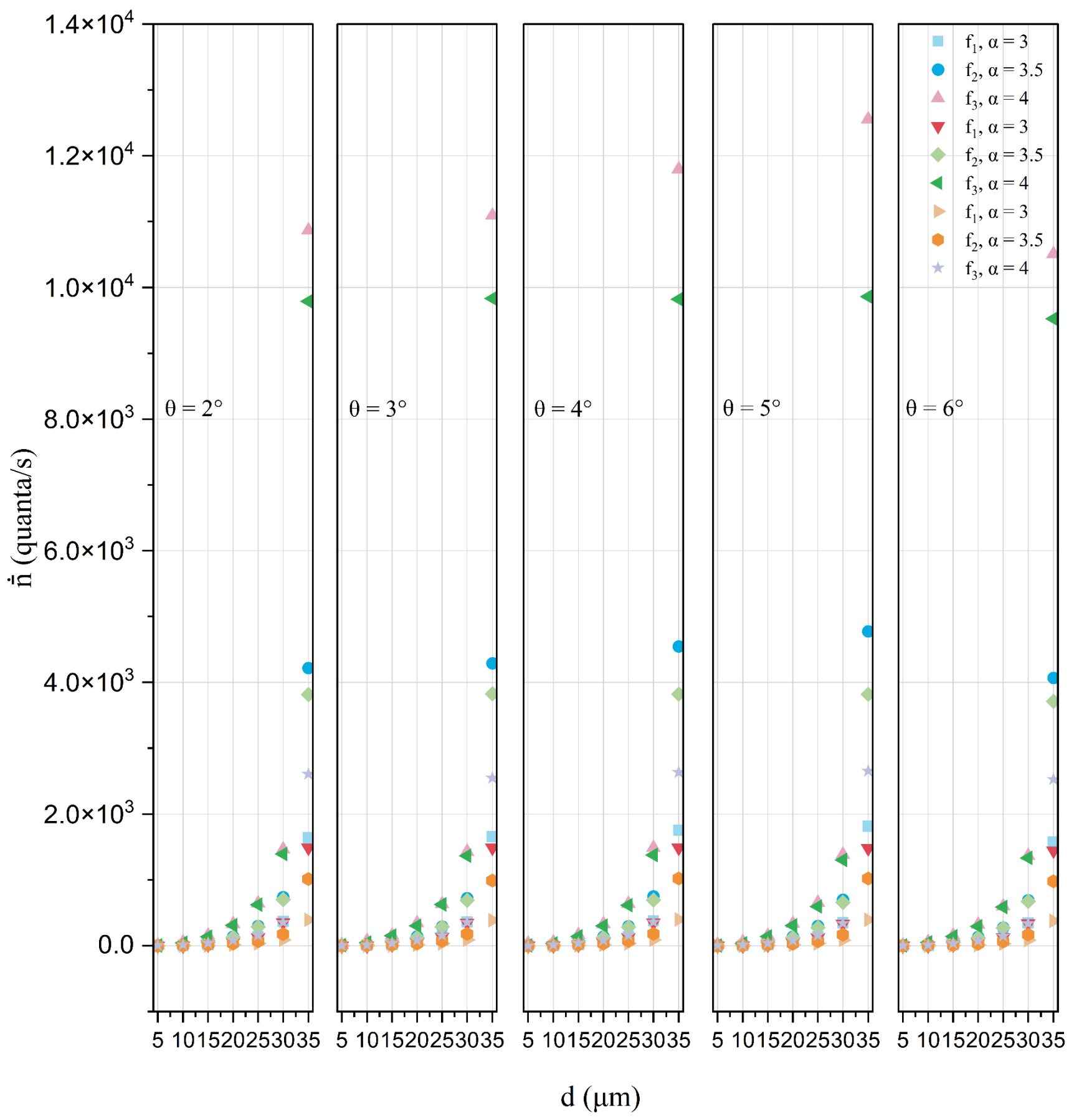

Figure 6. Calculated motional heating rates as a function of blade separation d for the five inclination angles, θ = 2°-6°. The curves represent the three normal modes for frequency-noise exponents α = 3, 3.5, and 4.

For example, at $\theta = 6°$, the heating rate of the first normal mode increases from approximately 4.70, 16.1, and 55.4 quanta $s^{-1}$ at d = 10 μm for $\alpha = 3$, 3.5, and 4, respectively, to approximately 352, 699, and 1385 quanta $s^{-1}$ at d = 30 μm. At d = 35 μm, the heating rate increases further because of the continued reduction in the secular frequencies. The higher-frequency third mode generally exhibits lower heating than the two lower-frequency modes because of the $1/\omega$ dependence of the heating rate.

The effect of the frequency exponent is also evident in Fig. 6. For a given blade separation and motional mode, increasing α from 3 to 4 produces a substantially higher predicted heating rate because the assumed electric-field noise increases more rapidly as the secular frequency decreases. This effect becomes strongest at the largest blade separations, where the secular frequencies are lowest. The five inclination angles exhibit similar overall trends, indicating that the blade separation and resulting secular frequency are the primary factors controlling the predicted heating, while the inclination angle produces comparatively smaller variations between the geometries.

### *3.4. Thermal response*

Fig. 7 shows the calculated temperature at the trapping position as a function of blade separation dd for the five electrode inclination angles, $\theta = 2°$-$6°$. All five curves exhibit nearly identical behaviour, indicating that the temperature at the trap position is only weakly dependent on the inclination angle over the investigated range. The temperature decreases systematically as the blade separation increases, from approximately 307.3 K at d = 5 μm to approximately 302.3 K at d = 35 μm for $\theta = 2°$-$6°$. The largest temperature reduction occurs between 5 and 10 μm, after which the temperature changes more gradually. Thus, the thermal response is relatively insensitive to the small variations in electrode inclination considered in this study.

The thermal behaviour was evaluated by imposing a blade temperature of 350 K and an external temperature of 300 K, with a heat-transfer coefficient of 10 $W\,m^{-2}\,K^{-1}$. The heat-transfer boundary condition represents the exchange of heat between the heated electrode surfaces and the surrounding environment. In this model, the resulting temperature at the trapping position is determined by the heat conduction through the electrode structure and the heat transfer from its surfaces to the surrounding environment.

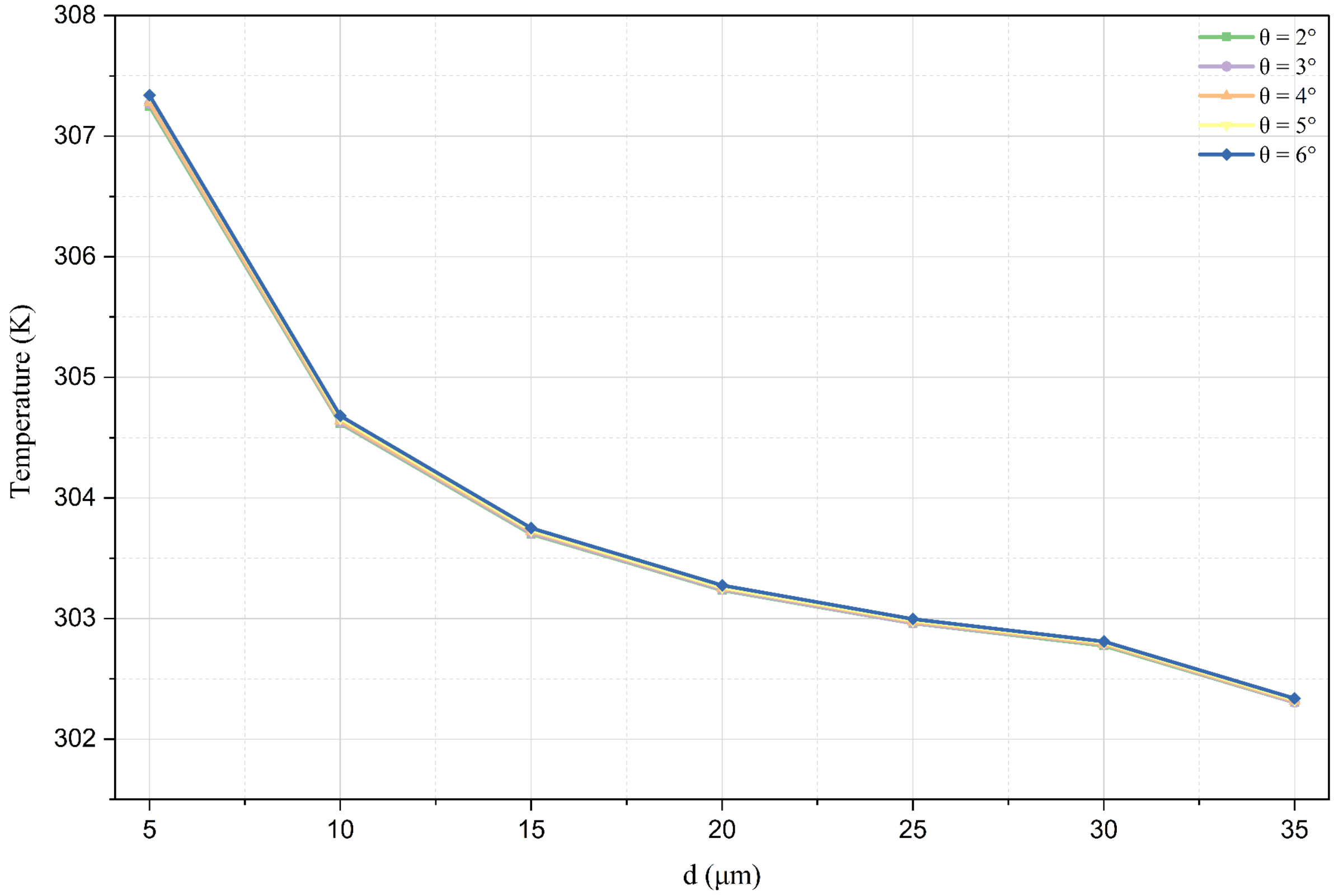

Figure 7. Calculated temperature at the trapping position as a function of blade separation d for electrode inclination angles θ = 2°-6°, with the blade temperature maintained at 350 K and the external temperature at 300 K.

The gradual decrease in temperature with increasing blade separation can be understood from the changing thermal proximity between the heated electrodes and the trapping position. At smaller separations, the trapping region is located closer to the heated electrode surfaces, resulting in a greater thermal influence at the ion position. As the blades are separated, the trapping position becomes farther from the heated surfaces and the temperature approaches the imposed external temperature of 300 K. For example, at θ = 6°, the calculated temperature decreases from approximately 307.3 K at d = 5 μm to 302.3 K at d = 35 μm, corresponding to a reduction of approximately 5 K.

### *3.5. RF stability analysis*

The RF stability of all five trap geometries was evaluated for blade separations of 5–35 μm using the matrix form of the Mathieu equation. The RF-potential Hessian obtained from COMSOL was used to construct the dimensionless RF matrix Q, retaining coupling between the three motional coordinates:

$$\frac{d^2\mathbf{u}}{d\tau^2} + [A - 2Q\cos cos\,(2\tau)\,]\mathbf{u} = 0 \text{ [25]},$$

where $\tau = \Omega t/2$. For the RF-only analysis, A = 0. The eigenvalues $q_i$ of Q were used to define

$$q_{max} = max_i\left|q_i\right| \text{ [25]}.$$

| d (μm) | θ = 2° | θ = 3° | θ = 4° | θ = 5° | θ = 6° |
|---|---|---|---|---|---|
| 5 | 426.607 | 448.907 | 515.470 | 458.017 | 489.932 |
| 10 | 40.779 | 40.958 | 41.152 | 43.024 | 41.587 |
| 15 | 10.737 | 10.745 | 10.679 | 10.852 | 10.940 |
| 20 | 4.099 | 4.134 | 4.132 | 4.161 | 4.292 |
| 25 | 1.902 | 1.894 | 1.898 | 1.905 | 1.966 |
| 30 | 0.880 | 0.878 | 0.873 | 0.899 | 0.905 |
| 35 | 0.522 | 0.523 | 0.518 | 0.527 | 0.534 |

Table : Maximum eigenvalue of the RF Mathieu matrix, $q_{max}$ , for the five trap inclinations at different blade separations.

The corresponding Floquet multipliers were then used to determine RF stability. At 15 kV and 28 MHz, configurations with d = 5–25 μm were unstable, whereas d=30 μm was stable but close to the stability boundary. At 30 μm, $q_{max}$ = 0.87–0.90, while at 35 μm, $q_{max}$ = 0.52–0.53, indicating a larger stability margin.

**4. Discussion**

The results in Figs. 3(a) and 3(b) show that the blade separation d is the main parameter controlling the strength of the RF confinement in the proposed inclined geometry. The almost complete overlap of the trap-depth curves for the five inclination angles in Fig. 3(a) indicates that changing θ from 2° to 6° has only a small effect on the overall trap depth. In contrast, increasing d produces a strong reduction in trap depth. This behaviour is consistent with the pseudopotential distribution shown in Fig. 3(b), where a clear minimum is formed at the trap centre and the surrounding higher-pseudopotential regions provide the confining barriers. The pseudopotential contours in Fig. 4 further show that this minimum is present in the xy, xz, and yz planes, confirming that the proposed inclined electrode arrangement produces three-dimensional confinement. The blade separation determines the trap's overall strength. As d increases, the ion is positioned farther from the electrodes and experiences a weaker RF field and field gradient, resulting in a shallower pseudopotential well.

The effect of this reduction in confinement is reflected directly in the secular-frequency behaviour shown in Fig. 5. All three normal-mode frequencies decrease strongly with increasing blade separation, particularly between 5 and 10 μm, and then decrease more gradually at larger separations. The similar trends for all five inclination angles further indicate that the change in d has a much stronger influence on the motional frequencies than the small change in θ. At small separations, the three modes are strongly separated, indicating that the ion experiences highly anisotropic confinement because of its close proximity to the electrode surfaces. With increasing d, the two lower-frequency modes become closely spaced, while the third mode remains approximately twice their average frequency. This behaviour is consistent with the pseudopotential contours in Fig. 4 and indicates that the local confinement becomes increasingly similar along two normal-mode directions as the ion moves farther from the electrodes, while the third direction retains stronger confinement. Thus, the frequency evolution in Fig. 5 provides a quantitative measure of the changes in pseudopotential curvature observed with increasing blade separation.

The change in secular frequency also provides the main link between the confinement results and the motional-heating behaviour in Fig. 6. Increasing the blade separation increases the ion-electrode distance, which reduces the distance-dependent electric-field noise according to the assumed $d^{-4}$ scaling. However, as shown in Fig. 5, the same increase in separation produces a substantial reduction in secular frequency. Since the heating rate is inversely proportional to the secular frequency, this reduction can compensate for or exceed the decrease in electric-field noise, resulting in an overall increase in the predicted heating rate. This explains the strong increase in heating observed in Fig. 6, particularly at d = 30-35 μm. The higher-frequency third mode generally shows lower heating than the two lower-frequency modes, which is consistent with the 1/ω dependence of the heating expression. The increasing difference between the curves for α = 3, 3.5, and 4 at larger separations also shows that the heating prediction becomes more sensitive to the assumed frequency dependence of the electric-field noise when the secular frequencies become small. Therefore, simply increasing the ion-electrode distance does not guarantee lower motional heating in the present design. Instead, the confinement strength and secular frequency must be considered together with the distance-dependent noise.

In comparison with the strong changes observed in Figs. 5, and 6, the thermal response shown in Fig. 7 is relatively weak. The five curves almost completely overlap, showing that the inclination angle has little influence on the temperature at the trapping position under the applied thermal boundary conditions. The temperature decreases from approximately 307.3 K at d = 5 μm to approximately 302.3 K at d = 35 μm. This decrease can be attributed to the increasing distance between the heated electrode surfaces and the trapping position. At smaller d, the trapping region is more strongly affected by the 350 K electrode temperature, whereas increasing d reduces this thermal influence and allows the temperature to approach the external temperature of 300 K. The small temperature variation across the investigated geometries indicates that, within the assumptions of the present thermal model, the thermal environment does not strongly distinguish between the five inclination angles. Thus, compared with RF confinement, secular frequency, and motional heating, the thermal response is not the dominant factor determining the choice of blade separation.

The RF stability analysis provides the final constraint on the design trends observed in Figs. 3-7. The smaller blade separations provide deeper pseudopotential wells and higher secular frequencies, as shown in Figs. 3 and 5, and also give lower predicted heating rates in Fig. 6. However, Table 1 shows that the configurations from d = 5 to 25 μm are RF unstable at 15 kV and 28 MHz. Therefore, the strong confinement obtained at these smaller separations cannot be considered a stable operating condition under the present RF parameters. At d = 30 μm, all five geometries become stable, but $q_{max}$ = 0.87-0.90 places them relatively close to the stability boundary. At d = 35 μm, $q_{max}$ =0.52-0.53 provides a substantially larger stability margin. This result changes the interpretation of the earlier confinement and heating trends: although reducing d is favourable from the viewpoint of trap depth, secular frequency, and predicted heating, it eventually leads to RF instability. Conversely, increasing d improves the RF stability but produces weaker confinement and higher predicted heating. Therefore, under the present operating conditions, the stable design region is approximately 25-35 μm, with near 25 μm offering stronger confinement but a smaller stability margin and 35 μm providing more robust RF stability at the cost of higher predicted heating. The inclination angle has a comparatively minor effect on these overall trends, suggesting that the blade separation is the more important parameter for further optimization of the proposed architecture.

## 5. Conclusion

This study investigated the confinement, motional dynamics, electric-field-noise-induced heating, thermal response, and RF stability of a three-dimensional inclined electrode ion-trap architecture using finite-element simulations. The results show that blade separation is the main parameter controlling the trap properties, while changing the inclination angle from 2° to 6° produces only a small effect on the overall behaviour. Increasing the blade separation reduces the trap depth and secular frequencies, while the predicted motional heating increases because the reduction in secular frequency outweighs the decrease in the distance-dependent electric-field noise. The thermal analysis shows only a small temperature increase at the trapping position under the applied thermal conditions.

The RF stability analysis provides the main constraint on the operating range. At 15 kV and 28 MHz, separations of 5–25 μm were found to be unstable, whereas stable RF confinement was obtained at near 25–35 μm. The 30 μm configuration provides stronger confinement and lower predicted heating but operates close to the stability boundary, while 35 μm provides a larger stability margin with weaker confinement and higher predicted heating. Thus, near 25–35 μm represents the most relevant operating range under the present conditions, with the final choice depending on the required balance between confinement strength, stability margin, and motional heating. These results provide a useful computational basis for further optimization and experimental investigation of the proposed inclined ion-trap architecture for quantum-information applications.

## References

[1] Pagano G, Adamczyk W and So V 2025 Fundamentals of trapped ions and quantum simulation of chemical dynamics arXiv:2505.20412

[2] Hong S, Lee M, Cheon H, Kim T and Cho D I D 2016 Guidelines for Designing Surface Ion Traps Using the Boundary Element Method Sensors 16 616

[3] Ospelkaus C, Warring U, Colombe Y, Brown K R, Amini J M, Leibfried D and Wineland D J 2011 Microwave quantum logic gates for trapped ions Nature 476 181

[4] Crain S, Mount E, Baek S and Kim J 2014 Individual addressing of trapped 171Yb+ ion qubits using a microelectromechanical systems-based beam steering system Appl. Phys. Lett. 105 181115

[5] Wild J P 1952 The Radio-Frequency Line Spectrum of Atomic Hydrogen and Its Applications in Astronomy Astrophys. J. 115 206

[6] Field G B 1958 Excitation of the Hydrogen 21-CM Line Proc. IRE 46 240

[7] Pritchard J R and Loeb A 2012 21 cm cosmology in the 21st century Rep. Prog. Phys. 75 086901

[8] Pi J, Liu X, Cao J, Wang P, Ou L, Gao E, Tu H, Zou M, Zhang X, Zhang J and Kim K 2026 Beyond-Ten-Hour Coherence in a Decoherence-Free Trapped-Ion Clock Qubit arXiv:2603.19631

[9] Blatt R and Wineland D J 2008 Entangled states of trapped atomic ions Nature 453 1008

[10] Udem Th, Holzwarth R and Hänsch T W 2002 Optical frequency metrology Nature 416 233

[11] Margolis H S, Barwood G P, Huang G, Klein H A, Lea S N, Szymaniec K and Gill P 2004 Hertz-level measurement of the optical clock frequency in a single 88Sr+ ion Science 306 1355

[12] Dawson P H (ed) 1976 Quadrupole Mass Spectrometry and Its Applications (Amsterdam: Elsevier)

[13] Cirac J I and Zoller P 1995 Quantum Computations with Cold Trapped Ions Phys. Rev. Lett. 74 4091

[14] Kielpinski D, Monroe C and Wineland D J 2002 Architecture for a large-scale ion-trap quantum computer Nature 417 709

[15] Hite D A, Colombe Y, Wilson A C, Allcock D T C, Leibfried D, Wineland D J and Pappas D P 2013 Surface science for improved ion traps MRS Bull. 38 826

[16] Bowler R, Gaebler J, Lin Y, Tan T R, Hanneke D, Jost J D, Home J P, Leibfried D and Wineland D J 2012 Coherent diabatic ion transport and separation in a multizone trap array Phys. Rev. Lett. 109 080502

[17] Walther A, Ziesel F, Ruster T, Dawkins S T, Ott K, Hettrich M, Singer K, Schmidt-Kaler F and Poschinger U 2012 Controlling Fast Transport of Cold Trapped Ions Phys. Rev. Lett. 109 080501

[18] Podoliak N, Takahashi H, Keller M and Horak P 2016 Comparative numerical studies of ion traps with integrated optical cavities Phys. Rev. Appl. 6 044008

[19] Stick D, Hensinger W K, Olmschenk S, Madsen M J, Schwab K and Monroe C 2006 Ion trap in a semiconductor chip Nat. Phys. 2 36

[20] Wang H, Xie Y, Tao Y, Wu W, Chen P and Chen T 2026 Mitigation of dielectric heating in surface-electrode ion traps Phys. Scr. 101 ae8338

[21] Autodesk Inc. 2021 Fusion 360 (San Francisco, CA: Autodesk Inc.)

[22] COMSOL AB 2021 COMSOL Multiphysics® v. 6.0 www.comsol.com (Stockholm: COMSOL AB)

[23] Descoeudres A, Ramsvik T, Calatroni S, Taborelli M and Wuensch W 2009 dc breakdown conditioning and breakdown rate of metals and metallic alloys under ultrahigh vacuum Phys. Rev. ST Accel. Beams 12 032001

[24] Abbasov T, Zibrov S and Sherstov I 2023 Surface-electrode ion trap development JETP Lett. 118 215

[25] Yoshimura B, Stork M, Dadic D, Campbell W C and Freericks J K 2015 Creation of two-dimensional Coulomb crystals of ions in oblate Paul traps for quantum simulations EPJ Quantum Technol. 2 2

[26] Leibfried D, Blatt R, Monroe C and Wineland D 2003 Quantum dynamics of single trapped ions Rev. Mod. Phys. 75 281

[27] Brownnutt M, Kumph M, Rabl P and Blatt R 2015 Ion-trap measurements of electric-field noise near surfaces Rev. Mod. Phys. 87 1419

[28] McKay K S, Hite D A, Kent P D, Kotler S, Leibfried D, Slichter D H, Wilson A C and Pappas D P 2021 Measurement of electric-field noise from interchangeable samples with a trapped-ion sensor Phys. Rev. A 104 052610

[29] Turchette Q A, Kielpinski D, King B E, Leibfried D, Meekhof D M, Myatt C J, Rowe M A, Sackett C A, Wood C S, Itano W M, Monroe C and Wineland D J 2000 Heating of trapped ions from the quantum ground state Phys. Rev. A 61 063418